\documentstyle[12pt]{article}
\def\be{\begin{equation}}
\def\ee{\end{equation}}
\def\bea{\begin{eqnarray}}
\def\eea{\end{eqnarray}}

\begin{document}

\begin{center}
{\Large{\bf PP-Wave Strings from Membrane 
and from String in the Spacetime with Two Time Directions}}                  
										 
\vskip .5cm   
{\large Davoud  Kamani}
\vskip .1cm
 {\it Institute for Studies in Theoretical Physics and
Mathematics (IPM)
\\  P.O.Box: 19395-5531, Tehran, Iran}\\
{\sl E-mail: kamani@theory.ipm.ac.ir}
\\
\end{center}

\begin{abstract} 

In this paper we obtain strings that propagate in the quantized pp-wave
backgrounds. We can obtain these strings from the solutions of membrane.
The other way is the propagation of a massless string in a spacetime
with two time dimensions. This string sweeps a worldvolume,
which enables us to obtain other strings in the quantized
pp-wave backgrounds in the spacetime with one time direction. The
associated algebras and Hamiltonians of these massive strings
will be studied.

\end{abstract} 
\vskip .5cm


\newpage

\section{Introduction}
The plane wave metric supported by a {\it Ramond}-{\it Ramond}
5-form background \cite{1}, provides examples of exactly solvable
string models \cite{2,3}. Note that, string theory in this background 
tests $AdS/CFT$ correspondence \cite{4}. This background by Penrose
limit \cite{5}, also is related to the $AdS_5 \times S^5$ background
\cite{6}.

Combining membrane theory \cite{7,8}
with string theory in the pp-wave background, we can find a structure
for the membrane \cite{9}.
In fact, an enlarged framework for studying strings is the compactified
membrane theory \cite{10}.
In this paper, from the solutions of the compacted membrane,
we shall obtain massive strings which propagate
in the pp-wave backgrounds. Some
properties of these strings such as algebra and Hamiltonian
will be studied.

Similar to the membrane theory, we shall also find strings in the
pp-wave backgrounds from a massless string that propagates in the
spacetime with two {\it time} directions. The worldvolume of this
massless string has one space and two time directions \cite{11,12}.
One of these two time coordinates
(both of the spacetime and of the string worldvolume) is compacted
on a circle. The resulted massive strings will be studied in detail.

This paper is organized as follows.
In section 2, relations between the massive strings with the membrane are
given. In section 3, from the solution of the membrane, we shall
obtain the solutions of the massive closed strings. In section 4,
the same will be done for the massive open strings. In section 5, by
compactifying the extra time coordinates of the spacetime and of the
string worldvolume, we shall obtain another solutions for the
massive closed and open strings.
\section{Membrane in terms of massive strings}

By choosing appropriate gauges \cite{8}, the membrane action can be
written as
\bea
S=-\frac{T_2}{2} \int d^2 \sigma
d \rho (\eta^{AB}\partial_AZ^I \partial_BZ^I),
\eea
where $I\in \{1,2,...,8\}$. The metric of the membrane worldvolume is
$\eta_{AB}={\rm diag}(-1,1,1)$ and $T_2$ is its tension.
Therefore, the equation of motion of the membrane is
\bea       
(-\partial_\tau^2 + \partial_\sigma^2+\partial_\rho^2 )Z^I(\tau, \sigma,
\rho)=0.
\eea

It is possible to expand $Z^I(\tau, \sigma, \rho)$ as in the following
\bea                        
Z^I(\tau,\sigma,\rho )=\sum^\infty_{n=-\infty} X^I_n(\tau,\sigma) q_n(\rho),
\eea
where $q_n(\rho)= \exp(\frac{in\rho}{R})$.
The ranges of $\sigma$ and $\rho$ are $0\leq \sigma \leq 2\pi\alpha' p^+$
and $0\leq \rho \leq 2\pi R$.
In fact, it is assumed that one of the spacetime and one of the membrane
directions are identified and wrapped on a circle with radius $R$.
The radius of compactification is
$R = \frac{1}{4 \pi^2 \alpha' T_2}$.
Since the membrane coordinate $Z^I(\tau,\sigma,\rho )$ is real, we should
have
\bea
X^{I\dagger}_n(\tau, \sigma)= X^I_{-n}(\tau,\sigma).
\eea

The equation of motion of a string with mass number $n$ is
\bea
(-\partial^2_\tau + \partial^2_\sigma -\mu^2_n)X^I_n (\tau,\sigma)=0,
\eea
where $\mu_n=\frac{n}{R}$ is the mass of the string.
This equation also can be obtained from the equations (2) and (3).

The Hamiltonian of this system of strings is
\bea
H= \frac{1}{4\pi \alpha'} \sum_{n\in Z}\int^{2\pi \alpha'p^+}_0 d \sigma
\bigg{(} \partial_\tau X^I_n\partial_\tau X^I_{-n} + \partial_\sigma X^I_n
\partial_\sigma X^I_{-n} + \mu^2_n X^I_n X^I_{-n} \bigg{)}.
\eea
This Hamiltonian describes infinite number of massive strings that interact
with each other. Strings with symmetric mass numbers
(i.e., $n$ and $-n$) couple to
each other. There is no coupling between strings with the same mass signs.
One interpretation is that, the membrane is a distribution of infinite number
of massive strings with quantized masses.

The Hamiltonian (6) can be written as $H=\sum_{n\in Z}H_n$. The
corresponding action also has the form $S=\sum_{n\in Z}S_n$. Each
action $S_n$ describes a massive string with the mass $\mu_n$. In
fact, after the gauge fixing \cite{2}, each $S_n$ is related to a plane
wave metric.

According to the equation (3) we have
\bea
X^I_n(\tau , \sigma) = \frac{1}{2\pi R} \int_{0}^{2\pi R} d \rho
\bigg{(}q_{-n}(\rho) Z^I(\tau, \sigma,\rho)\bigg{)}.
\eea
Therefore, integration of the membrane coordinates by the weight
$q_{-n}(\rho)$ over the worldvolume coordinate $\rho$, produces
massive string coordinates.
\section{Massive closed strings}

A solution of the equation of motion of the membrane, i.e., equation (2),
can be written as
\bea
Z^I(\tau, \sigma, \rho) = z^I(\tau ,\rho) +
\frac{i\sqrt{2\alpha'}}{2\alpha' p^+}\sum_{(l,k) \neq (0,0)}
\bigg{[}\frac{1}{\Omega_{lk}}\exp[-i(\Omega_{lk}\tau-\mu_k \rho)]
\bigg{(} A^I_{lk}e^{i{\bar l}\sigma} +
{\tilde A}^I_{lk}e^{-i{\bar l}\sigma} \bigg{)} \bigg{]},
\eea
where the zero modes are given by the function
$z^I(\tau ,\rho)$ as in the following
\bea
z^I(\tau ,\rho) = \sum_{k\in Z}\bigg{[} q_k(\rho)\bigg{(}
x^I_k \cos(\mu_k \tau) + \frac{p^I_k}{p^+}\frac{\sin(\mu_k \tau)}{\mu_k}
\bigg{)}\bigg{]} .
\eea
Also $\Omega_{lk}$ and ${\bar l}$ are
\bea
\Omega_{lk} ={\rm sgn}(l)\sqrt{\bigg{(}\frac{l}{\alpha' p^+}\bigg{)}^2 +
\bigg{(}\frac{k}{R}\bigg{)}^2}\;
\;\;\;\;,\;\;\;\;{\bar l}= \frac{l}{\alpha'p^+}.
\eea
The worldvolume frequencies $\{\Omega_{lk}\}$ imply that the indices
``$l$'' and ``$k$'', in the oscillating part of the solution (8),
simultaneously can not be zero.
The Fourier coefficients have been chosen so that $Z^I$ is real coordinate.
Therefore, we should have
\bea
&~&x^{I\dagger}_k = x^I_{-k} ,
\nonumber\\
&~&p^{I\dagger}_k = p^I_{-k} ,
\nonumber\\
&~& A^{I\dagger}_{lk} = A^I_{-l,-k} ,
\nonumber\\
&~& {\tilde A}^{I\dagger}_{lk} = {\tilde A}^I_{-l,-k} .
\eea
Furthermore, the algebra of the membrane modes are
\bea
[A^I_{lk}\; , \;A^J_{l'k'}]
=[{\tilde A}^I_{lk}\; , \;{\tilde A}^J_{l'k'}]
= \alpha' p^+ \Omega_{lk}\delta^{IJ} \delta_{l+l',0}\delta_{k+k',0}.
\eea
\bea
[x^I_k \;,\;p^J_{k'}] = i \delta^{IJ} \delta_{k+k',0}.
\eea

Combining the solution (8) with the equation (7), we obtain
the closed string coordinates with the mass number $n$,
\bea
X^I_n(\tau, \sigma) = x^I_n \cos(\mu_n \tau) + \frac{p^I_n}{p^+}
\frac{\sin(\mu_n \tau)}{\mu_n}+
\frac{i\sqrt{2\alpha'}}{2\alpha' p^+} \sum_{(l,n) \neq (0,0)}
\bigg{[}\frac{1}{\Omega_{ln}}e^{-i\Omega_{ln}\tau}
\bigg{(} A^I_{ln}e^{i{\bar l}\sigma} +
{\tilde A}^I_{ln}e^{-i{\bar l}\sigma} \bigg{)} \bigg{]}.
\eea
This string coordinate satisfies the condition (4).
Note that $A^I_{-l,-n}$ for positive indices $l$ and $n$ is
creation operator, while $A^I_{ln}$ is annihilation operator. Also
$A^I_{-l,n}$ with respect to the index $l$ is creation operator and with
respect to the index $n$ is annihilation operator. Similar interpretation
also holds for $A^I_{l,-n}$.

The equation (6) gives the Hamiltonian of the closed strings as in the
following
\bea
H=\frac{1}{2 p^+} \sum_{n\in Z}\bigg{(}p^I_{-n}p^I_n+
m^2_n x^I_{-n}x^I_n \bigg{)}+
\frac{1}{2\alpha' p^+}\sum_{n \in Z}
\sum_{(l,n) \neq (0,0)}\bigg{(} A^I_{-l,-n}A^I_{ln}+{\tilde A}^I_{-l,-n}
{\tilde A}^I_{ln} \bigg{)}.
\eea
Now define the new oscillators $a^I_{ln}$ and ${\tilde a}^I_{ln}$ as
\bea
&~& a^I_{ln} = \frac{1}{\sqrt{\alpha' p^+ |\Omega_{ln}|}} A^I_{ln},
\nonumber\\
&~& {\tilde a}^I_{ln} = \frac{1}{\sqrt{\alpha' p^+ |\Omega_{ln}|}}
{\tilde A}^I_{ln}.
\eea
These oscillators have the algebras
\bea
[a^I_{ln}\; , \;a^J_{l'n'}] = [{\tilde a}^I_{ln}\; , \;{\tilde a}^J_{l'n'}]
={\rm sgn}(l)\delta^{IJ}\delta_{l+l',0} \delta_{n+n',0}.
\eea
Therefore, the oscillating part of the Hamiltonian (15) takes the form
\bea
H_{\rm osc} = \sum^\infty_{n=1}\bigg{[}|\Omega_{0n}|\bigg{(}
a^I_{0,-n}a^I_{0n} + {\tilde a}^I_{0,-n} {\tilde a}^I_{0n}\bigg{)}\bigg{]}+ 
\sum_{n\in Z}\sum^\infty_{l=1}\bigg{[}|\Omega_{ln}|\bigg{(}
a^I_{-l,-n}a^I_{ln} + {\tilde a}^I_{-l,-n} {\tilde a}^I_{ln}\bigg{)}\bigg{]}+ 
C + {\tilde C}.
\eea
The constants $C$ and ${\tilde C}$ come from the normal ordering of the
Hamiltonian. After regularization, they are
\bea
C = {\tilde C}= -\frac{1}{3}\bigg{(}\frac{1}{R}+\frac{1}{\alpha' p^+}
\bigg{)} + 8 \sum^{\infty}_{n=1}\sum^{\infty}_{l=1} \Omega_{ln}.
\eea
In $H_{\rm osc}$, 
the first summation is normal ordered. Also the second summation
with respect to the index $l$ is normal ordered.
\section{Massive open strings}

The procedure of closed string also can be applied for open string. That
is, for the equation (2), we can consider the solution
\bea
Z^I(\tau, \sigma, \rho) = z^I(\tau , \rho) + 
\frac{i\sqrt{2\alpha'}}{2\alpha' p^+}\sum_{(l,k) \neq (0,0)}
\bigg{[}\frac{1}{\Omega'_{lk}} A^I_{lk}\exp [-i(\Omega'_{lk}\tau-
\mu_k\rho)] \cos \bigg{(}\frac{l\sigma}{2\alpha'p^+}\bigg{)} \bigg{]},
\eea
where $z^I(\tau , \rho)$ has been given by the equation (9), and
$\Omega'_{lk}$ is
\bea
\Omega'_{lk} = {\rm sgn}(l)\sqrt{\bigg{(} \frac{l}{2\alpha' p^+}\bigg{)}^2 +
\bigg{(}\frac{k}{R}\bigg{)}^2}.
\eea
Since the coordinate $Z^I(\tau, \sigma, \rho)$ is real the Fourier modes
should satisfies the conditions
\bea
&~&x^{I\dagger}_k = x^I_{-k} ,
\nonumber\\
&~&p^{I\dagger}_k = p^I_{-k} ,
\nonumber\\
&~& A^{I\dagger}_{lk} = A^I_{-l,-k}.
\eea

The open string coordinates resulted from the membrane coordinates (20),
satisfy the Neumann boundary condition. For satisfying the Dirichlet boundary
condition, it is sufficient to drop $z^I(\tau , \rho)$ and change
$\cos (\frac{l\sigma}{2\alpha' p^+})$ to
$-i\sin(\frac{l\sigma}{2\alpha' p^+})$.

The equations (7) and (20) give
the coordinates of the open string with the mass number $n$, 
\bea
X^i_n(\tau, \sigma) = x^i_n \cos(\mu_n \tau) + \frac{p^i_n}{p^+}
\frac{\sin(\mu_n \tau)}{\mu_n}+
\frac{i\sqrt{2\alpha'}}{2\alpha' p^+}\sum_{(l,n) \neq (0,0)}
\bigg{[}\frac{1}{\Omega'_{ln}} A^i_{ln}e^{-i\Omega'_{ln}\tau}
\cos \bigg{(}\frac{l\sigma}{2\alpha' p^+}\bigg{)} \bigg{]},
\eea
for the Neumann directions, and
\bea
X^a_n(\tau, \sigma) =
\frac{\sqrt{2\alpha'}}{2\alpha' p^+}\sum_{(l,n) \neq (0,0)}
\bigg{[}\frac{1}{\Omega'_{ln}} A^a_{ln}e^{-i\Omega'_{ln}\tau}
\sin \bigg{(}\frac{l\sigma}{2\alpha' p^+}\bigg{)} \bigg{]},
\eea
for the Dirichlet directions. The coordinates (23) and (24) satisfy
the condition (4).
For the massless string i.e., $n=0$,
these are the usual open string coordinates.

The Hamiltonian of this system of open strings is sum of the Dirichlet
and Neumann Hamiltonians. Therefore, we have
\bea
H=\frac{1}{2 p^+} \sum_{n\in Z}\bigg{(}p^i_{-n}p^i_n+
m^2_n x^i_{-n}x^i_n \bigg{)}+
\frac{1}{4\alpha' p^+}\sum_{n \in Z}\sum_{(l,n)
\neq (0,0)} ( A^I_{-l,-n}A^I_{ln}).
\eea

The algebras of the open string modes are equation (13) and
\bea
[A^I_{ln}\; , \;A^{J}_{l'n'}]
= 2\alpha' p^+ \Omega'_{ln}\delta^{IJ} \delta_{l+l',0}\delta_{n+n',0}\;\;,
\;\;\;\;I,J\in\{i,a\}.
\eea
Define the oscillators $\{a^I_{ln}\}$ as in the following
\bea
a^I_{ln} = \frac{1}{\sqrt{2\alpha' p^+ |\Omega'_{ln}|}} A^{I}_{ln}.
\eea
Then the algebra of these oscillators becomes
\bea
[a^I_{ln}\; , \;a^J_{l'n'}]={\rm sgn}(l)\delta^{IJ}\delta_{l+l',0}
\delta_{n+n',0}\;,\;\;\;\;I,J\in\{i,a\}.
\eea
In terms of these oscillators and after normal ordering, the
oscillating part of the Hamiltonian (25) takes the form
\bea
H_{\rm osc}= \sum^\infty_{n=1}\bigg{(}|\Omega'_{0n}|
a^I_{0,-n}a^I_{0n}\bigg{)}+ 
\sum_{n \in Z}\sum^\infty_{l =1}\bigg{(}|\Omega'_{ln}|
a^I_{-l,-n}a^I_{ln}\bigg{)}+ C',
\eea
where the normal ordering constant $C'$ is
\bea
C' = -\frac{1}{3}\bigg{(}\frac{1}{R}+\frac{1}{2\alpha' p^+}\bigg{)} +
8 \sum^{\infty}_{n=1}\sum^{\infty}_{l=1} \Omega'_{ln}.
\eea
\section{Strings in the pp-wave backgrounds from
string in the spacetime with two time directions}

Some efforts have been devoted to imagining a world with more than one
{\it time} dimension. There are many possibilities for the numbers of
space dimension, time dimension, and also for the space and time
dimensions of the extended objects, that propagate in the spacetime
\cite{11,12}. For example, it is possible to have a membrane that its
worldvolume has two space and one time dimensions.
This membrane propagates in
a spacetime with ten space and one time dimensions.
There is another interesting
possibility, i.e., a string that its worldvolume contains one space
and two times. This string lives in a spacetime with nine spaces and
two times. So, both the $(10,1;2,1)$ and the $(9,2;1,2)$ cases yield
$SO(3,2) \times SO(8)$ as the bosonic group \cite{11}.

In addition to $Z^0$, let the direction $Z^{10}$ also be a time coordinate
of the spacetime. Therefore, the action of a massless
string in this spacetime is
\bea
{\bar S}=-\frac{T}{2}\int d\sigma d^2\tau (\eta_{\mu\nu}
\eta^{AB}\partial_AZ^\mu \partial_BZ^\nu -1),
\eea
where $\mu ,\nu \in \{0,1,...,10\}$. The parameters $\tau$ and $\tau'$
are time coordinates and $\sigma$ is space coordinate of the 
string worldvolume. The metrics of the spacetime and
the worldvolume are $\eta_{\mu\nu} = {\rm diag}(-1,1,1,...,1,-1)$ and
$\eta_{AB} = {\rm diag}(-1,-1,1)$.
The constant $T$ is tension of the string in this spacetime. It is not equal
to $\frac{1}{2\pi \alpha'}$ which is string tension in the usual spacetime
(i.e., spacetime with one time direction).
By the identification $Z^{10} = \tau'$ and choosing the light-cone gauge,
the action (31) takes the form
\bea
{\bar S}=-\frac{T}{2}\int d^2\tau\int^{2\pi \alpha'p^+}_0
d\sigma (\eta^{AB}\partial_AZ^I
\partial_BZ^I).
\eea
The equation of motion of string, extracted from this action, is
\bea
(-\partial^2_\tau -\partial^2_{\tau'}+\partial^2_\sigma )
Z^I(\tau , \tau' ;\sigma)=0.
\eea
In addition, for the closed string there is the condition
\bea
Z^I(\tau , \tau' ;\sigma+2\pi \alpha' p^+) = Z^I(\tau , \tau' ;\sigma ).
\eea
The open string also should satisfy the conditions
\bea
(\partial_\sigma Z^I)_{\sigma_0} = 0,
\eea
for the Neumann directions, and
\bea
&~& (\partial_\tau Z^I)_{\sigma_0} = 0,
\nonumber\\
&~& (\partial_{\tau'} Z^I)_{\sigma_0} = 0,
\eea
for the Dirichlet directions. The constant $\sigma_0 = 0, 2\pi \alpha' p^+$
shows the ends of the open string.

Assume that the extra time direction $Z^{10}$ to be compact on
a circle with the radius $R$.
Since the functions $\{ q_n(\tau')= \exp{(\frac{in\tau'}{R})}\}$ are
periodic we can write
\bea                        
Z^I(\tau,\tau';\sigma )=\sum^\infty_{n=-\infty}X^I_n(\tau,\sigma)
q_n(\tau').
\eea
Again the reality of $Z^I$ leads to the condition (4) for the
coordinate $X^I_n{(\tau,\sigma)}$. According to the equation
(37) we have
\bea
X^I_n(\tau , \sigma) = \frac{1}{2\pi R} \int_{0}^{2\pi R} d \tau'
\bigg{(}q_{-n}(\tau') Z^I(\tau, \tau';\sigma)\bigg{)}.
\eea

The equation (33) gives the equation of motion of the string
coordinate $X^I_n(\tau , \sigma)$ as
\bea
(-\partial^2_\tau + \partial^2_\sigma +\mu^2_n)X^I_n (\tau,\sigma)=0.
\eea
Also the string tension $T$ depends on the radius of
compactification, i.e.,
\bea
T = \frac{1}{4 \pi^2 \alpha' R}.
\eea
According to the equation (37), the action (32) becomes
\bea
{\bar S}=-\frac{1}{4\pi \alpha'}\sum_{n\in Z}
\int d\tau \int^{2\pi \alpha'p^+}_0 d \sigma
\bigg{(}\eta^{ab}\partial_a X^I_n \partial_b X^I_{-n}
- \mu^2_n X^I_n X^I_{-n}\bigg{)}.
\eea
In fact, the equation of motion extracted from this action, is the
equation (39). This action leads to the following Hamiltonian
\bea
{\bar H} = \frac{1}{4\pi \alpha'} \sum_{n\in Z}\int^{2\pi
\alpha'p^+}_0 d \sigma
\bigg{(} \partial_\tau X^I_n\partial_\tau X^I_{-n} + \partial_\sigma X^I_n
\partial_\sigma X^I_{-n} - \mu^2_n X^I_n X^I_{-n} \bigg{)}.
\eea

The action (41) and the associated Hamiltonian (42) can be written as
${\bar S}=\sum_{n\in Z} {\bar S}_n$ and
${\bar H}=\sum_{n\in Z} {\bar H}_n$. Each light-cone action ${\bar S}_n$
corresponds to the plane wave metric
\bea
d{{\bar s}_n}^2 = 2 dX^+dX^- +\mu^2_n \sum^8_{I=1} X^I_n X^I_{-n}
(dX^+)^2 + \sum^8_{I=1} dX^I_n dX^I_{-n}\;.
\eea
For achieving to this metric, the role of the coordinates
$X^0$ and $X^9$ should be changed. In other words, in the usual plane
wave metric, we should apply the exchanges $X^0 \rightarrow \pm iX^9$
and $X^9 \rightarrow \pm iX^0$. Therefore, the light-cone coordinates
$X^{\pm} = \frac{1}{\sqrt{2}}(X^0 \pm X^9)$ transform as
$X^+ \rightarrow \pm iX^+$ and $X^- \rightarrow \mp iX^-$.

{\bf Closed strings in pp-waves}

For a closed string the solution of the equation of motion (33) is
\bea
Z^I(\tau, \tau'; \sigma) = z^I(\tau,\tau')
+ \frac{i\sqrt{2\alpha'}}{2\alpha' p^+}
\sum_{k\in Z}\sum_{|l|\geq l_k+1} 
\bigg{[}\frac{1}{\omega_{lk}}\exp[-i(\omega_{lk}\tau-\mu_k\tau')]
\bigg{(} A^I_{lk}e^{i{\bar l}\sigma} +
{\tilde A}^I_{lk}e^{-i{\bar l}\sigma} \bigg{)} \bigg{]},
\eea
where the zero modes are given by the function
\bea
z^I(\tau ,\tau') = \sum_{k\in Z}\bigg{[} q_k(\tau')\bigg{(}
x^I_k \cosh(\mu_k \tau) + \frac{p^I_k}{p^+}\frac{\sinh(\mu_k \tau)}{\mu_k}
\bigg{)}\bigg{]}.
\eea
Also $\omega_{lk}$, ${\bar l}$ and $l_k$ are 
\bea
\omega_{lk} ={\rm sgn}(l)\sqrt{\bigg{(}\frac{l}{\alpha'p^+}
\bigg{)}^2 - \bigg{(}\frac{k}{R}\bigg{)}^2}\;
\;\;\;\;,\;\;\;\;{\bar l}= \frac{l}{\alpha'p^+}\;\;\;,
\;\;\;l_k = \bigg{[} \frac{\alpha' p^+}{R}|k| \bigg{]} .
\eea
In fact, the number $l_k$ is integer part of
$\frac{\alpha' p^+}{R}|k|$. Reality
of the coordinate (44) implies that the Fourier coefficients
should obey the conditions (11).

The coordinate of the closed string with the mass number $n$,
in the usual spacetime is
\bea
X^I_n(\tau, \sigma) = x^I_n \cosh(\mu_n \tau) + \frac{p^I_n}{p^+}
\frac{\sinh(\mu_n \tau)}{\mu_n}+
\frac{i\sqrt{2\alpha'}}{2\alpha' p^+} \sum_{|l|\geq l_n+1}
\bigg{[}\frac{1}{\omega_{ln}}e^{-i\omega_{ln}\tau}
\bigg{(} A^I_{ln}e^{i{\bar l}\sigma} +
{\tilde A}^I_{ln}e^{-i{\bar l}\sigma} \bigg{)} \bigg{]}.
\eea
This coordinate satisfies the equation of motion (39) and the condition (4).
The algebra of the modes of this coordinate, has been given by the
equations (12) and (13).

The equations (42) and (47) give the Hamiltonian of the closed strings as
\bea
H=\frac{1}{2 p^+} \sum_{n\in Z}\bigg{(}p^I_{-n}p^I_n
- m^2_n x^I_{-n}x^I_n \bigg{)}+
\frac{1}{2\alpha' p^+}\sum_{n \in Z}
\sum_{|l|\geq l_n+1}\bigg{(} A^I_{-l,-n}A^I_{ln}+{\tilde A}^I_{-l,-n}
{\tilde A}^I_{ln} \bigg{)}.
\eea
Using the definition (16), with $\omega_{ln}$ instead of $\Omega_{ln}$,
the oscillating part of this Hamiltonian becomes
\bea
H_{\rm osc}=\sum_{n\in Z}\sum^\infty_{l=l_n+1}\bigg{[}|\omega_{ln}|\bigg{(}
a^I_{-l,-n}a^I_{ln} + {\tilde a}^I_{-l,-n} {\tilde a}^I_{ln}\bigg{)}\bigg{]}+ 
A + {\tilde A}.
\eea
The normal ordering constants $A$ and ${\tilde A}$ come from the algebra
(17) and they are
\bea
A = {\tilde A}= -\frac{1}{3\alpha' p^+}
+ 8 \sum^{\infty}_{n=1}\sum^{\infty}_{l=l_n+1} \omega_{ln}.
\eea

{\bf Open strings in pp-waves}

The equation of motion (33) gives the following solutions for
the open strings 
\bea
Z^i(\tau,\tau'; \sigma) = z^i(\tau ,\tau') + 
\frac{i\sqrt{2\alpha'}}{2\alpha' p^+}\sum_{k\in Z} \sum_{|l|\geq l_k+1}
\bigg{[}\frac{1}{\omega'_{lk}} A^i_{lk}\exp [-i(\omega'_{lk}\tau-
\mu_k\tau')] \cos \bigg{(}\frac{l\sigma}{2\alpha' p^+}\bigg{)} \bigg{]},
\eea
for the Neumann boundary condition, where $z^i(\tau ,\tau')$ has been given
by the equation (45) and $\omega'_{lk}$ and $l_k$ are
\bea
\omega'_{lk} = {\rm sgn}(l)\sqrt{\bigg{(}\frac{l}{2\alpha' p^+} \bigg{)}^2 -
\bigg{(}\frac{k}{R}\bigg{)}^2}
\;\;\;\;,\;\;\;\; l_k=\bigg{[} \frac{2\alpha' p^+}{R}|k| \bigg{]}.
\eea
For the Dirichlet boundary condition we have
\bea
Z^a(\tau, \tau';\sigma) = \frac{\sqrt{2\alpha'}}{2\alpha' p^+}
\sum_{k\in Z} \sum_{|l|\geq l_k+1}
\bigg{[}\frac{1}{\omega'_{lk}} A^a_{lk}\exp [-i(\omega'_{lk}\tau-
\mu_k\tau')] \sin \bigg{(}\frac{l\sigma}{2\alpha' p^+}\bigg{)} \bigg{]}.
\eea
The reality of the coordinates $Z^i$ and $Z^a$ leads to the conditions
(22) for the open string modes.

The open string coordinates in the usual spacetime are
\bea
X^i_n(\tau, \sigma) =x^i_n \cosh(\mu_n \tau) + \frac{p^i_n}{p^+}
\frac{\sinh(\mu_n \tau)}{\mu_n}+
\frac{i\sqrt{2\alpha'}}{2\alpha' p^+} \sum_{|l|\geq l_n+1}
\bigg{[}\frac{1}{\omega'_{ln}} A^i_{ln}e^{-i\omega'_{ln}\tau}
\cos \bigg{(}\frac{l\sigma}{2\alpha' p^+}\bigg{)} \bigg{]},
\eea
for the Neumann directions, and
\bea
X^a_n(\tau, \sigma) =
\frac{\sqrt{2\alpha'}}{2\alpha' p^+}\sum_{|l|\geq l_n+1}
\bigg{[}\frac{1}{\omega'_{ln}} A^a_{ln}e^{-i\omega'_{ln}\tau}
\sin \bigg{(}\frac{l\sigma}{2\alpha' p^+}\bigg{)} \bigg{]},
\eea
for the Dirichlet directions. These solutions also satisfy the condition (4).

The Hamiltonian of this system of open strings is
\bea
H'=\frac{1}{2 p^+} \sum_{n\in Z}\bigg{(}p^i_{-n}p^i_n
- m^2_n x^i_{-n}x^i_n \bigg{)}+
\frac{1}{4\alpha' p^+}\sum_{n \in Z}
\sum_{|l|\geq l_n+1} ( A^I_{-l,-n}A^I_{ln}) .
\eea
Apply the definition (27), with $\omega'_{ln}$ instead of $\Omega'_{ln}$,
we obtain the oscillating part of $H'$ as
\bea
H'_{\rm osc} =
\sum_{n \in Z}\sum^\infty_{l =l_n+1}\bigg{(}|\omega'_{ln}|
a^I_{-l,-n}a^I_{ln}\bigg{)}+ A'.
\eea
The algebra (28) gives the normal ordering constant $A'$,
\bea
A' = -\frac{1}{6\alpha' p^+} + 8 \sum^{\infty}_{n=1}
\sum^{\infty}_{l=l_n+1} \omega'_{ln}.
\eea
Note that the equation (13) gives the algebra of $x^I_n$ and $p^I_n$ for
both closed strings and the Neumann part of open strings.
\section{Conclusions}

In terms of the membrane solutions, we obtained the solutions of the
massive closed and open strings, the
algebras of the modes and the associated Hamiltonians.
In fact, the algebra of the strings modes in
the quantized pp-wave backgrounds
is the same as the algebra of the membrane modes.
The normal ordering constants of the Hamiltonians
depend on both the string length $2\pi \alpha' p^+$ and the radius of
compactification of the membrane.

We also obtained strings in the quantized pp-wave backgrounds from a 
massless string that propagates in the spacetime with two time directions.
One of these time coordinates is compacted on circle.
The resulted massive strings are open or closed. The open string
coordinates satisfy the Neumann or Dirichlet boundary conditions,
as the usual case. Reality of the massless string coordinates
projects out some of the oscillators
from the solutions of the resulted massive strings and therefore,
from the associated Hamiltonians.

\end{document}